\documentclass[12pt]{article}
\usepackage{epsfig}
\usepackage{amsmath}
\usepackage{amssymb}

\def\dfrac#1#2{\displaystyle \frac{#1}{#2}}
\def\dsum{\mathop{\displaystyle \sum }}

\newcommand{\msbar}{\overline{\mbox{\scriptsize MS}}}
\newcommand{\MSbar}{\overline{\mbox{MS}}}
\newcommand{\nn}{\nonumber}
\newcommand{\simge}{\ \lower-
1.2pt\vbox{\hbox{\rlap{$>$}\lower5pt \vbox{\hbox{$\sim$}}}}\ }
\newcommand{\calo}{{\cal O}}

\newcommand{\be}{\begin{equation}}
\newcommand{\ee}{\end{equation}}
\newcommand{\ba}{\begin{eqnarray}}
\newcommand{\ea}{\end{eqnarray}}
\newcommand{\brr}{\begin{array}}
\newcommand{\err}{\end{array}}

\begin{document}
\thispagestyle{empty} 
\begin{flushright}
\begin{tabular}{l}
{\tt CPT-2004/P.027} \\
{\tt FTUV-04-0519}\\
{\tt IFIC/04-22}\\
{\tt RM3-TH/04-12}\\
{\tt ROME1-1374/2004}\\
{\tt ROM2F/2004/16}
\end{tabular}
\end{flushright}
\begin{center}

\vskip 1.2cm
{\Large \bf Non-perturbative renormalization of lattice }\\ 
\vskip 0.2cm
{\Large \bf operators in coordinate space}\\

\vskip 0.9cm 
{\sc V.~Gim\'enez$^a$, L.~Giusti$^b$, S.~Guerriero$^c$, V.~Lubicz$^c$, 
G. Martinelli$^{d,e}$}
\vskip 0.2cm
{\sc S.~Petrarca$^{d,e}$, J.~Reyes$^{d}$, B.~Taglienti$^{e}$, E.~Trevigne$^{f}$}

\vskip 0.7 cm
{\sl
$^a$ Dep. de F{\'\i}sica Th\`eorica and IFIC, Univ. de Val\`encia, \\
Dr. Moliner 50, E-46100, Burjassot, Val\`encia, Spain \\
\vskip 0.1cm
$^b$ Centre de Physique Th\'eorique, CNRS Luminy, Case 907, F-13288 Marseille,
France \\
\vskip 0.1cm
$^c$  Dip. di Fisica, Universit\`a di Roma Tre and INFN, Sezione di Roma III, \\
Via della Vasca Navale 84, I-00146 Roma, Italy \\
\vskip 0.1cm
$^d$ Dip. di Fisica, Universit\`a di Roma ``La Sapienza'', P.le A. Moro 2, 
I-00185 Rome, Italy \\
\vskip 0.1cm
$^e$ INFN, Sezione di Roma, P.le A. Moro 2, I-00185 Rome, Italy \\
\vskip 0.1cm
$^f$ Dip. di Fisica, Universit\`a di Roma ``{\it Tor Vergata}'' and INFN, 
Sezione di Roma 2 \\ Via della Ricerca Scientifica, I-00133 Roma, Italy\\
}
\vskip1.cm
\end{center}

\begin{abstract}
We present the first numerical implementation of a non-perturbative 
renormalization method for lattice operators, based on the study of correlation
functions in coordinate space at short Euclidean distance. The method is applied
to compute the renormalization constants of bilinear quark operators for the 
non-perturbative $O(a)$-improved Wilson action in the quenched approximation. 
The matching with perturbative schemes, such as $\MSbar$, is computed at the 
next-to-leading order in continuum perturbation theory. A feasibility study of 
this technique with Neuberger fermions is also presented.
\end{abstract}
\vskip 2.2cm

\renewcommand{\thefootnote}{\arabic{footnote}}
\vspace*{-1.5cm}

\newpage
\setcounter{page}{1}
\setcounter{equation}{0}
\setcounter{footnote}{0}
\section{Introduction}

Correlation functions of quantum-field operators are computed non-perturbatively
on the lattice by Monte Carlo simulations. From their long-distance behaviour,
in QCD, non-perturbative features of the underlying theory, such as the hadron 
spectrum and matrix elements, can be extracted. On the other hand, their 
behaviour at short distance is expected to be controlled by perturbation theory
and by the operator product expansion (OPE). As a result, relevant ultraviolet 
informations of the theory, such as the range of applicability of the OPE, the 
values of the Wilson coefficients and the renormalization constants of composite
operators, can in principle be obtained by a comparison of the perturbative
formul\ae~with the numerical results at short distance. This is the basic idea 
behind the non-perturbative renormalization techniques proposed in the last 
decade and widely used in present simulations. With the RI/MOM method, 
renormalization conditions are imposed on quark and gluon Green functions 
computed non-perturbatively in momentum space, in a fixed gauge, with given 
off-shell external states of large 
virtuality~\cite{Martinelli:1994ty}-\cite{Gattringer:2004iv}. 
In the Schr\"odinger functional (SF) approach, the renormalization 
conditions are imposed in coordinate space at a given finite physical distance 
on suitable gauge invariant correlation functions in finite volume with SF 
boundary conditions \cite{Luscher:1992an}-\cite{Capitani:1998mq}. The 
step-scaling technique~\cite{Luscher:1991wu} can then be used to convert 
the renormalization constants to their renormalization group invariant 
definitions. An advantage of these methods is that 
the conversion to more popular continuum schemes,
such as the $\MSbar$ scheme, can be implemented by performing a calculation only
in continuum perturbation theory, by comparing renormalized correlation 
functions at short distances computed in dimensional regularization in the two 
schemes. In this way, more tedious calculations with lattice perturbation theory
are completely avoided.

In this paper we present the first numerical implementation of a 
non-perturbative renormalization method based on the study of lattice 
correlation functions at short Euclidean distances in coordinate 
space~\cite{Martinelli:1997zc}. We call this approach the ``{\it X-space}" 
scheme. Preliminary results were presented in 
Refs.~\cite{Becirevic:2002yv,Gimenez:2003rt}. We have computed numerically the 
two-point functions of all dimension three bilinear quark operators by 
discretizing the gluons {\em a l\'a} Wilson and the fermions with the 
non-perturbatively $O(a)$-improved Wilson action. A feasibility study of this 
technique with Neuberger fermions is also presented. The multiplicatively 
renormalization constants of lattice bilinear operators are evaluated 
non-perturbatively by imposing renormalization conditions directly in x-space at
short distance. The condition $x_0\gg a$, where $a$ is the lattice spacing and 
$x_0$ is the renormalization point, has to be satisfied in order to keep 
discretization errors under control. On the other hand, the matching of the 
renormalization constants to the $\MSbar$ scheme (or any other continuum scheme)
can be computed in continuum perturbation theory when $x_0\ll \Lambda^{-1}_{\rm 
QCD}$. In this study we show that for $a \lesssim 0.05$~fm (i.e. $1/a \gtrsim 4$
GeV) it is possible to find a region on the same lattice in which perturbation 
theory can be applied and discretization effects are still under control. The 
existence of the window $a \lesssim x_0 \lesssim \Lambda^{-1}_{\rm QCD}$ 
requires however rather fine lattices, large volumes and therefore expensive 
simulations. Alternatively, one can appeal to a step-scaling technique 
analogous to the one proposed in Ref.~\cite{Luscher:1991wu}. The X-space renormalization method 
involves only gauge-invariant correlation functions among local operators at 
finite physical distance, and can be easily applied to any fermion 
discretization. It can be very powerful for the evaluation of the 
renormalization constants of composite operators, such as the four-fermion 
operators relevant for the phenomenology of hadronic weak decays. The matching 
to the $\MSbar$ scheme can be performed by using only continuum perturbation 
theory and the method is very simple to implement.

The paper is organized as follows: in sec.~\ref{sec:PT} we summarize the 
relevant formul\ae~in continuum perturbation theory and define the 
renormalization conditions which we will use in coordinate space; in 
sec.~\ref{sec:results} we discuss the numerical results and in 
sec.~\ref{sec:concl} we present our conclusions. 
 
\section{X-space renormalization and perturbation theory}\label{sec:PT}

In this section we define the X-space renormalization scheme for bilinear quark 
operators and provide the perturbative expressions, at the next-to-leading order
(NLO), needed to convert the results to any other continuum renormalization 
scheme, such as the $\MSbar$ scheme\footnote{All formulas presented in this 
section are obtained in the infinite-volume limit. When correlation functions
in small volumes are considered, their perturbative expressions may need to be 
modified according to the boundary conditions used \cite{Luscher:1982uv,
Luscher:1982ma}}.

We consider the correlation functions of flavor non-singlet bilinear quark 
operators of the form
\be
\langle O_\Gamma(x) O_\Gamma(0) \rangle \: ,
\ee
where 
\be
O_\Gamma(x) = \bar \psi(x) \Gamma \psi(x)
\ee
with $O_\Gamma=\{S,P,V_\mu,A_\mu\,T_{\mu\nu}\}$ for $\Gamma=\{1,\,\gamma_5,\,
\gamma_\mu,\,\gamma_\mu\gamma_5,\,\frac{1}{2}[\gamma_\mu,\gamma_\nu]\,\}$ 
respectively and with flavor indices omitted. 

Following \cite{Martinelli:1997zc}, we impose non-perturbatively, in x-space and
in the chiral limit, the renormalization conditions 
\be\label{eq:rin}
\lim_{a\to 0} \langle O^{\rm X}_\Gamma(x) O^{\rm X}_\Gamma(0) \rangle\Big|
_{x^2=x_0^2} = \langle O_\Gamma(x_0) O_\Gamma(0) \rangle^{\rm free}_{\rm cont}
\ee
where the renormalized operator is 
$O^{\rm X}_\Gamma(x,x_0) = Z^{\rm X}_{\Gamma}(x_0)\, O_\Gamma(x)$ and $x_0$ is 
the renormalization point. The renormalization condition (\ref{eq:rin}) defines
the X-space scheme. Note that $x_{0}$ must satisfy the condition $a\ll x_{0}
\ll\Lambda_{QCD}^{-1}$ to keep non-perturbative and discretization effects under
control.

In order to illustrate the procedure and get the expressions needed to convert 
to the more popular continuum schemes, we have computed the correlation 
functions, at two loop in na\"\i ve dimensional regularization (NDR), in the 
massless case. The results, in Euclidean space, read
\ba
&&\hspace{-1cm}\langle S(x) S(0) \rangle =  \langle P(x) P(0) \rangle  =
\frac{N_c}{\pi^4\,{(x^2)}^3} \left\{1 +\frac{2\,\alpha_s}{4 \pi}\left(
\frac{4}{\hat \epsilon} + \frac{4}{3} + 8\,\gamma_E - \frac{\gamma^{(0)}_S}{2} 
\ln(\frac{\mu^2 x^2}{4}) \right)\right\} 
\nn \\
&&\hspace{-1cm}\langle V_\mu(x) V_\nu(0) \rangle = \langle A_\mu(x) A_\nu(0) 
\rangle = -\frac{2\,N_c}{ \pi^4\,{(x^2)}^3}\left(\frac{1}{2}\delta_{\mu\nu}-
\frac{x_\mu x_\nu}{x^2}\right) \left\{1 + 4 \frac{\alpha_s}{4 \pi}\right\} 
\label{eq:cc_ndr}\\
&&\hspace{-1cm}\langle T_{\mu\nu}(x) T_{\rho\sigma}(0) \rangle = -\frac{2\,N_c}
{\pi^4 \,{(x^2)}^3} \left(\frac{1}{2}T^{(1)}_{\mu\nu\rho\sigma}-T^{(2)}
_{\mu\nu\rho\sigma}\right)
\nn\\
&&\phantom{\hspace{-1cm}\langle T_{\mu\nu}(x) T_{\rho\sigma}(0) \rangle =}
\left\{1 +\frac{2\,\alpha_s}{4 \pi}\left(-\frac{4}{3\,\hat \epsilon} + 4 -
\frac{8}{3}\,\gamma_E - \frac{\gamma^{(0)}_T}{2} \ln(\frac{\mu^2 x^2}{4})
\right)\right\}
\nn
\ea
where
\ba
T^{(1)}_{\mu\nu\rho\sigma}&=&\delta_{\mu\rho}\delta_{\nu\sigma}-\delta_
{\mu\sigma}\delta_{\nu\rho}\nn\\
T^{(2)}_{\mu\nu\rho\sigma}&=&\frac{x_\mu x_\rho}{x^2}\delta_{\nu\sigma}-
\frac{x_\mu x_\sigma}{x^2}\delta_{\nu\rho}-\frac{x_\nu x_\rho}{x^2}\delta
_{\mu\sigma} +\frac{x_\nu x_\sigma}{x^2}\delta_{\mu\rho} \,.
\ea
In these expressions $\alpha_s$ is the strong coupling constant, $1/\hat\epsilon
 = 1/\epsilon -\ln(4\pi) -\gamma_E$ (we define $d=4-2\epsilon$ the space-time
dimension) and the LO anomalous dimensions are $\gamma^{(0)}_S=-8$, $\gamma^
{(0)}_T=8/3$, and $\gamma_V^{(0)}=0$, the latter due to the conservation of the 
vector current. This conservation also determines the tensor structure of the 
vector current correlator. Note that, as expected, the leading short
distance behaviour of the correlation functions in Eq.~(\ref{eq:cc_ndr}) is 
governed by $(x^2)^{-3}$. For the scalar and vector correlators 
the results in Eqs.~(\ref{eq:cc_ndr}) agree with previous 
computations~\cite{Chetyrkin:1996sr,Gorishnii:1990vf,Surguladze:1990tg}.

By imposing the renormalization conditions ~(\ref{eq:rin}) to the results in
Eq.~(\ref{eq:cc_ndr}), we obtain the correlation functions for the renormalized 
operators in the X-space scheme: 
\ba
\langle S^{\rm X}(x,x_0) S^{\rm X}(0,x_0) \rangle &= & \frac{N_c}
{\pi^4\,{(x^2)}^3} K_S^{\rm X}\Big(x,x_0\Big) 
\nn\\
\langle V^{\rm X}_\mu(x,x_0) V^{\rm X}_\nu(0,x_0) \rangle & = & -\frac{2\,N_c}
{ \pi^4\,{(x^2)}^3}\left(\frac{1}{2}\delta_{\mu\nu}-\frac{x_\mu x_\nu}{x^2}
\right) K_V^{\rm X}\Big(x,x_0\Big)
\label{eq:cc}\\ 
\langle T^{\rm X}(x,x_0)_{\mu\nu} T^{\rm X}(0,x_0)_{\rho\sigma} 
\rangle &= & -\frac{2\,N_c}{\pi^4\,{(x^2)}^3} \left(\frac{1}{2}T^{(1)}
_{\mu\nu\rho\sigma}-T^{(2)}_{\mu\nu\rho\sigma}\right) K_T^{\rm X}\Big(x,x_0
\Big) \nn
\ea
with 
\be
K_\Gamma^{\rm X}\Big(x,x_0\Big) = 1 -\gamma_\Gamma^{(0)} \frac{\alpha_s}{4 \pi} 
\ln\Big(\frac{x^2}{x_0^2}\Big) \, .
\ee

In the $\MSbar$ scheme, on the other hand, the correlation functions of the 
renormalized composite operators, $O^{\msbar}_\Gamma(x,\mu) = Z^{\msbar}
_{\Gamma}(\mu)\, O_\Gamma(x)$, are obtained by subtracting the pole $1/\hat
\epsilon$ on the r.h.s of Eq.~(\ref{eq:cc_ndr}). The relations between the 
$\MSbar$ and the X-space scheme are thus the following
\ba
K_S^{\overline{MS}}\Big(x,\mu\Big) & = & \left\{1 +\frac{2\alpha_s}{4\pi}
\left[4\ln\left(\frac{\mu^2 x_0^2}{4}\right)+8\gamma_E+\frac{4}{3}\right] 
\right\} K^{\rm X}_S\Big(x,x_0\Big)\nn\\
K_V^{\overline{MS}}\Big(x,\mu\Big) & =  & \Big(1+4\frac{\alpha_s}{4\pi}\Big) 
K^{\rm X}_V\Big(x,x_0\Big)\\
K_T^{\overline{MS}}\Big(x,\mu\Big) & = & \left\{1 +\frac{2\alpha_s}{4\pi}
\left[-\frac{4}{3}\ln\left( \frac{\mu^2 x_0^2}{4}\right)-\frac{8}{3} \gamma_E+
4 \right] \right\}K^{\rm X}_T\Big(x,x_0\Big)\, .\nn
\ea
This also implies that the renormalization constants in the two schemes are 
related by
\ba
\frac{Z^{\overline{\rm MS}}_S(\mu)}{Z^{\rm X}_S(x_0)} & = &  1 +\frac{\alpha_s}
{4 \pi}\left[4\ln{\left(\frac{\mu^2 x_0^2}{4}\right)}+8\gamma_E+\frac{4}{3}
\right]\; ,
\nn\\ 
\frac{Z^{\overline{\rm MS}}_V(\mu)}{Z^{\rm X}_V(x_0)} & = & 1 + 2\frac{\alpha_s}
{4 \pi}, 
\label{eq:zz}\\
\frac{Z^{\overline{\rm MS}}_T(\mu)}{Z^{\rm X}_T(x_0)} & = & 1 +\frac{\alpha_s}
{4\pi}\left[-\frac{4}{3}\ln\left(\frac{\mu^2 x_0^2}{4}\right)-\frac{8}{3} 
\gamma_E+4 \right] \, .\nn
\ea
The renormalization condition (\ref{eq:rin}) does not satisfy
the vector and axial vector Ward Identities. At the NLO this can be 
easily seen  
by noticing that in the $\overline{\rm MS}$ scheme, 
which preserves them, $K_V^{\overline{MS}}(x_0,\mu)$
has a finite term proportional to $\alpha_s$. In
the X-space scheme this contribution is included in the renormalization
constant and therefore $Z^{\overline{\rm MS}}_V/Z^{\rm X}_V \neq 1$, i.e.
the Ward Identities are broken.
They can be recovered by using continuum perturbation theory, which 
for the vector correlator is known up to four loops 
\cite{Gorishnii:1990vf,Surguladze:1990tg}, or 
non-perturbatively by matching the result for  $Z_V$ in the X-space  scheme 
with the Ward Identity determination~\cite{Bochicchio:1985xa}.

We conclude this section by recalling the expression for the renormalization
group evolution of the renormalization constants at the NLO in $\alpha_s$:
\be
Z^{\msbar}_\Gamma(\mu') =  \frac{c^{\overline{\rm MS}}_{\Gamma}(\mu')}
{c^{\overline{\rm MS}}_{\Gamma}(\mu)} Z^{\msbar}_\Gamma(\mu)
\ee 
where 
\be
c_{\Gamma}^{\overline{\rm MS}} \left(\mu \right)  = \alpha_s 
(\mu)^{\frac{\gamma^{(0)}_{\Gamma}}{2\beta_0}}\left\{ 1 +\frac{\alpha_s}{4\pi } 
\left(\frac{{\gamma }^{(1)}_{\Gamma}}{2\beta_0}-
\frac{\beta_1}{\beta_0} \frac{\gamma^{(0)}_{\Gamma}}{2\beta_0}\right) \right\}\;
\label{eq:calfa}
\ee
with $\alpha_s$ defined in $\overline{\rm MS}$ scheme. In the quenched theory, 
i.e. with $N_f=0$, the first two coefficients of the expansion of the 
$\beta$-function are $\beta_0=11$ and $\beta_1=102$, while the two-loop 
anomalous dimensions in the $\overline{\rm MS}$ scheme are $\gamma^{(1)}_S= 
-404/3$, $\gamma^{(1)}_V=0$ and $\gamma^{(1)}_T = 
724/9$~\cite{Chetyrkin:1997dh,Vermaseren:1997fq,Gracey:2000am}. In our study,
the matching between X-space and $\MSbar$ schemes has been performed at a scale 
$\mu
\sim 1/x_0$. With this choice, logs in Eqs.~(\ref{eq:zz}) are small and need not
to be resummed. In contrast, larger logs enter the evolution from the scale 
$\mu\sim 1/x_0$ to the conventional scale $\mu=2$ GeV at which our final results
are quoted. For this reason, the evolution function in Eq.~(\ref{eq:calfa}) has 
been resummed by using the renormalization group equations at the NLO.

\section{Numerical results}\label{sec:results}

In this section we provide the numerical details of our computation and present 
the results obtained non-perturbatively for the renormalization constants of the
bilinear quark operators.

\begin{figure}
\begin{center}
\begin{tabular}{c}
\epsfxsize7.5cm\epsffile{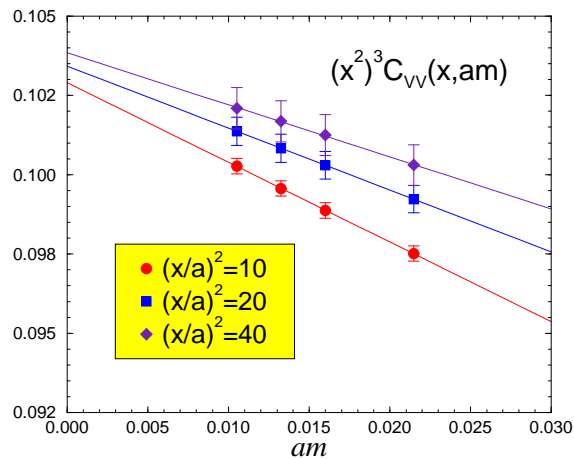} 
\end{tabular}
\caption{\sl\small The vector correlator $C_{VV}(x)$ for different values of
$x^2$ as a function of the quark mass. The line represents the result of the 
linear extrapolation to the chiral limit. \label{fig:chiral}}
\vspace*{-0.5cm}
\end{center}
\end{figure}

We generated a sample of $180$ gauge configurations in the quenched
approximation with the standard SU(3) Wilson gluonic action at $\beta=6.45$ 
($a\sim 0.048$~fm) and $V=32^3 \times 70$. For these configurations we evaluated
fermion propagators with the non-perturbatively $O(a)$-improved Wilson action 
for hopping parameter values $\kappa$=0.1349, 0.1351, 0.1352, 0.1353. We 
computed the two-point flavor non-singlet correlation functions 
\be
\label{eq:latcor}
\begin{array}{ll}
C_{SS}(x) = \phantom{\hspace{-0.5cm}\dsum\limits_\mu }
\langle S(x)S(0) \rangle \quad , & 
\hspace{-1.0cm} C_{PP}(x) = \langle P(x)P(0) \rangle \quad ,\\
C_{VV}(x) = \dsum\limits_{\mu} \langle V_\mu(x) V_\mu(0) \rangle \quad ,& 
\hspace{-1.0cm} C_{AA}(x) = \dsum\limits_\mu \langle A_\mu(x) A_\mu(0) \rangle 
\quad ,\\ C_{TT}(x) = \dsum\limits_{\mu,\nu,\rho} 
\left(\dfrac{1}{6}\, \delta_{\mu\nu}-\dfrac{1}{3}\, \frac{x_\mu x_\nu}{x^2}
\right)\langle T_{\mu\rho}(x)T_{\nu\rho}(0) \rangle &
\end{array}\ee
of local bilinear operators in the standard way and estimated the statistical 
errors by a jackknife procedure. These functions have been averaged over points 
which are equivalent under hypercubic rotations. In the range of $x^2$ we have 
studied, our data show a mild mass dependence, and a linear or quadratic
extrapolations to the chiral limit give compatible results within the errors. 
In the following we will show the results linearly extrapolated to the chiral 
limit. An example of this extrapolation, in the case of the vector correlator, 
is illustrated in Fig.~\ref{fig:chiral}.

\begin{figure}
\begin{tabular}{cc}
\epsfxsize7.5cm\epsffile{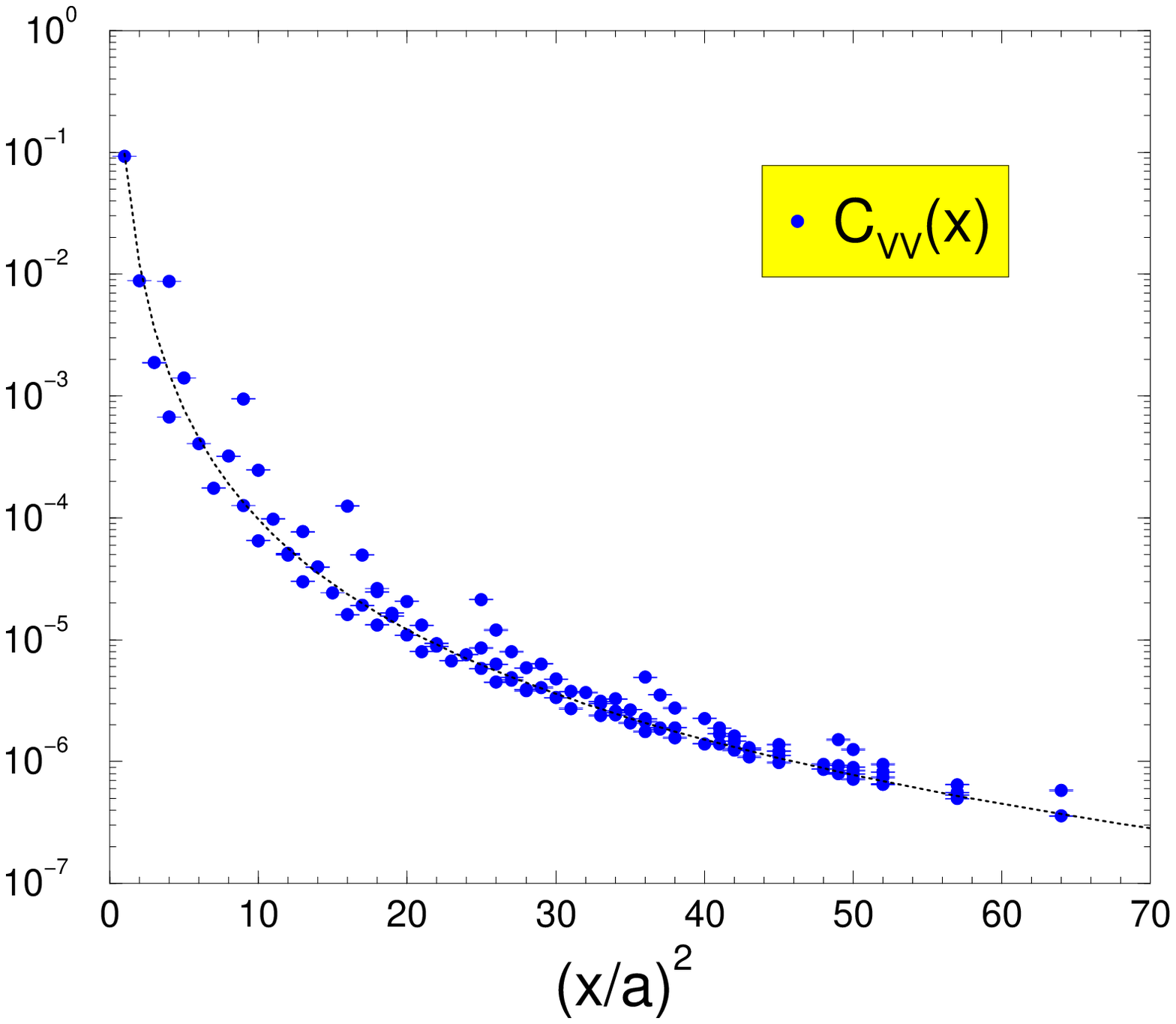} &
\hspace*{0.2cm}
\epsfxsize7.5cm\epsffile{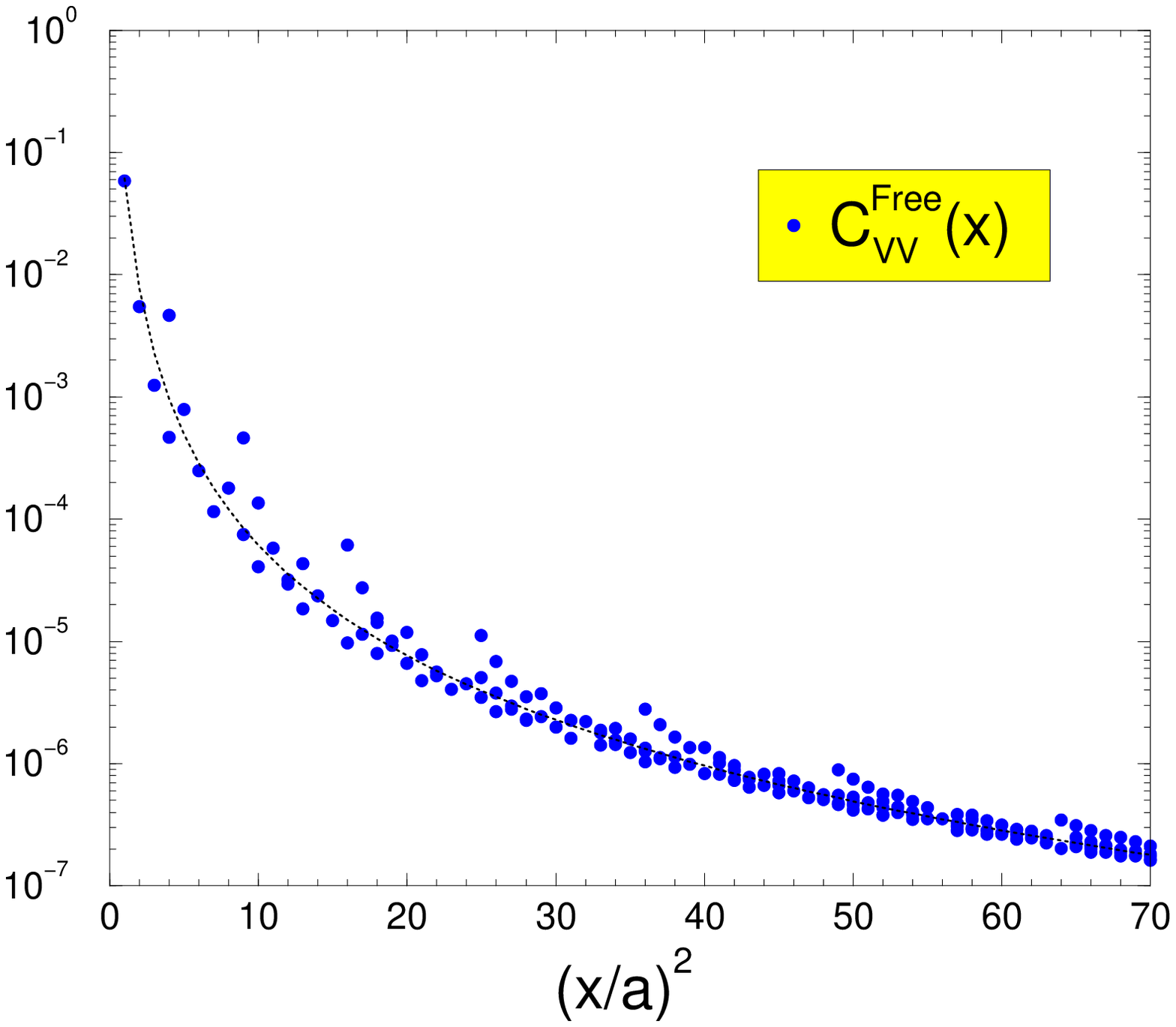} \\
\end{tabular}
\caption{\sl\small $C_{VV}(x)$ in the interacting (left) and in the free (right)
theory. In the first case, the dotted curve is a one parameter fit showing the 
${(x^2)}^{-3}$ behaviour. In the free case the curve represents the prediction 
of the free theory in the continuum limit\label{fig:vv_chi_int}}
\vspace*{-0.5cm}
\end{figure}
\begin{figure}
\begin{center}
\begin{tabular}{c}
\epsfxsize7.5cm\epsffile{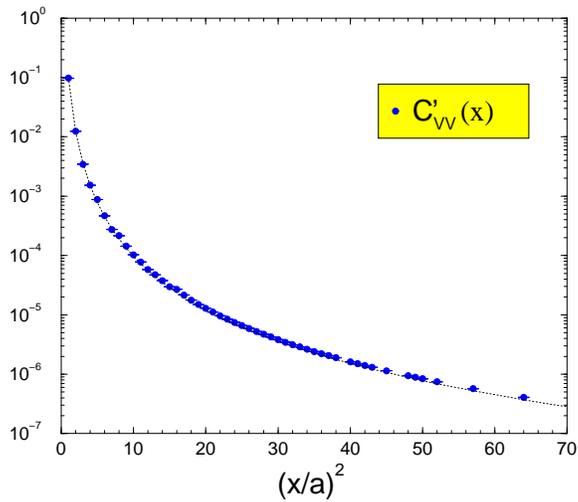} 
\end{tabular}
\caption{\sl\small The corrected vector correlator $C^\prime_{VV}(x)$. The 
dotted curve is a one parameter fit showing the ${(x^2)}^{-3}$ 
behaviour.\label{fig:vv_chi_corr}}
\vspace*{-0.5cm}
\end{center}
\end{figure}
In Fig.~\ref{fig:vv_chi_int} (left) we show the correlation function $C_{VV}(x)$
as a function of $x^2$ and the corresponding one parameter fit to $A\,{(x^2)}^
{-3}$. The na\"\i ve expected behaviour is clearly satisfied, even if a large 
scattering of the points due to lattice artifacts is visible, particularly in 
the short distance region. Lorentz invariance requires the correlator in the 
continuum limit to be a function of $x^2$ only. The lattice data presented in
Fig.~\ref{fig:vv_chi_int} (left) show instead that the results for $C_{VV}(x)$ 
computed at points which correspond to the same value of $x^2$ are often quite 
separated. To clarify the origin of these effects, we studied the correlators 
in the free theory as a function of volume, lattice spacing and quark masses. 
The free theory prediction for the lattice correlation function $C_{VV}(x)$, in 
infinite volume and in the chiral limit, is shown as an example in 
Fig.~\ref{fig:vv_chi_int} (right). The spread of the data observed in the 
interacting case turns out to be well reproduced in the free theory, at fixed 
volume and lattice spacing. For values of $x^2$ in the perturbative region, 
finite volume effects are found to be negligible in the range of masses we use. 
On the other hand, the spread of the data in the free case is considerable 
reduced by decreasing the lattice spacing. This suggests that the dominant 
contributions due to lattice artifacts comes from discretization effects.

In order to reduce discretization effects in the interacting case, we define
``corrected" correlation functions 
\be\label{riscala}
C_{\Gamma\Gamma}^{\prime}(x)=\frac{C_{\Gamma\Gamma}(x)}{\Delta_{\Gamma\Gamma}(x)
}
\ee
where $\Delta_{\Gamma\Gamma}(x)$ is the ratio of free correlator on the lattice 
over the continuum one, computed in infinite volume and in the chiral limit,
\be
\Delta_{\Gamma\Gamma}(x)=\frac
{\langle O_{\Gamma}(x)O_{\Gamma}(0) \rangle_{\rm lat}^{\rm free}}
{\langle  O_{\Gamma}(x)O_{\Gamma}(0) \rangle_{\rm cont}^{\rm free}}\,.
\end{equation}
By construction, $\Delta_{\Gamma\Gamma}(x)$ is equal to unit up to 
discretization effects. The results for the corrected function $C_{VV}^{\prime}
(x)$ are shown in Fig.~\ref{fig:vv_chi_corr}. These results, as well as those 
used in the following analysis, have been also averaged over points which 
correspond to the same $x^2$. The comparison between $C_{VV}(x)$ and $C_{VV}
^{\prime}(x)$ shows that, once tree-level discretization effects are removed, 
the spread of the data is greatly reduced. This analysis further supports the 
interpretation that the spread in the interacting theory is due to 
discretization effects. Similar conclusions apply also to other correlators.

\begin{figure}
\begin{tabular}{cc}
\epsfxsize7.5cm\epsffile{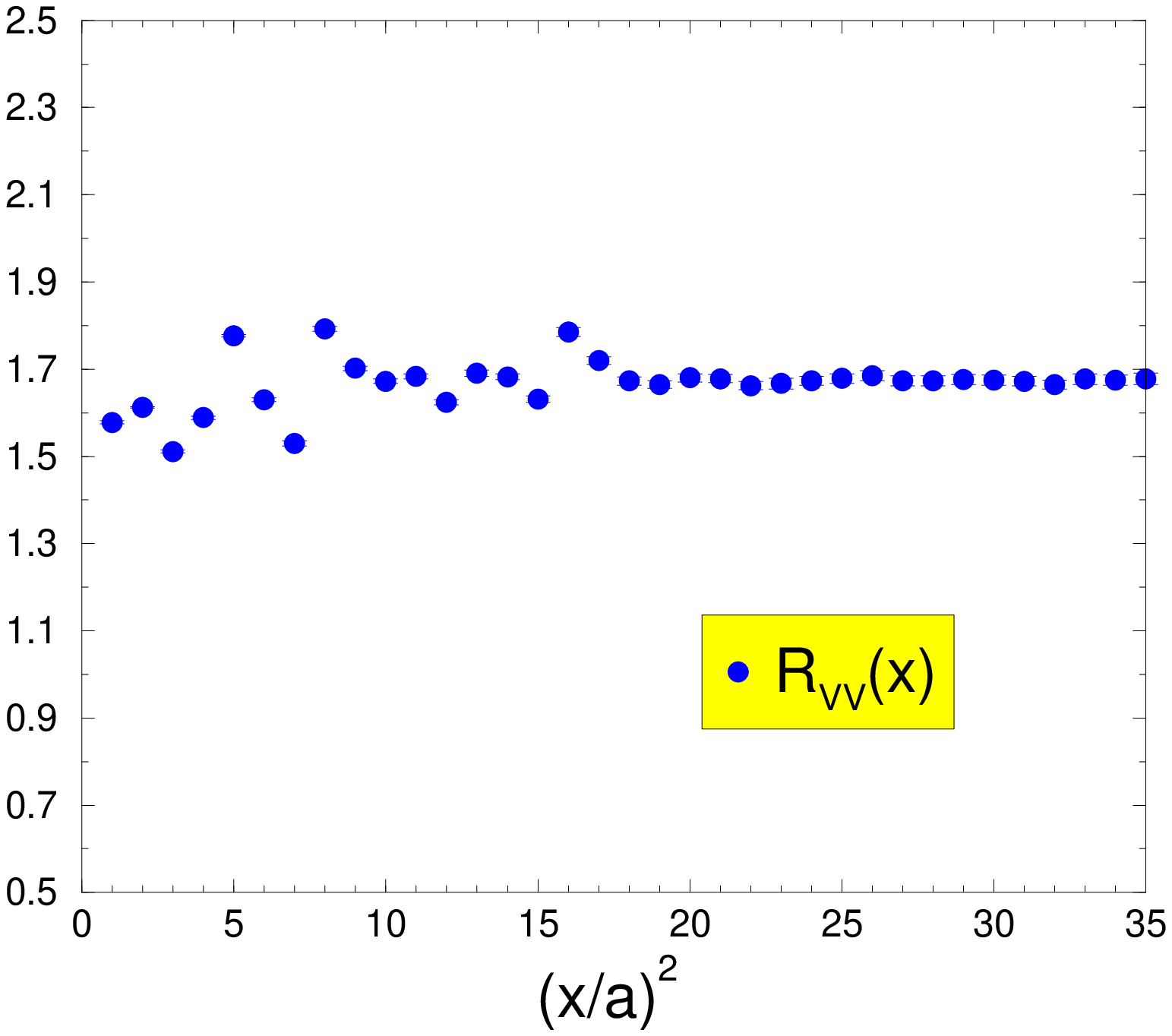} &
\hspace*{0.2cm}
\epsfxsize7.5cm\epsffile{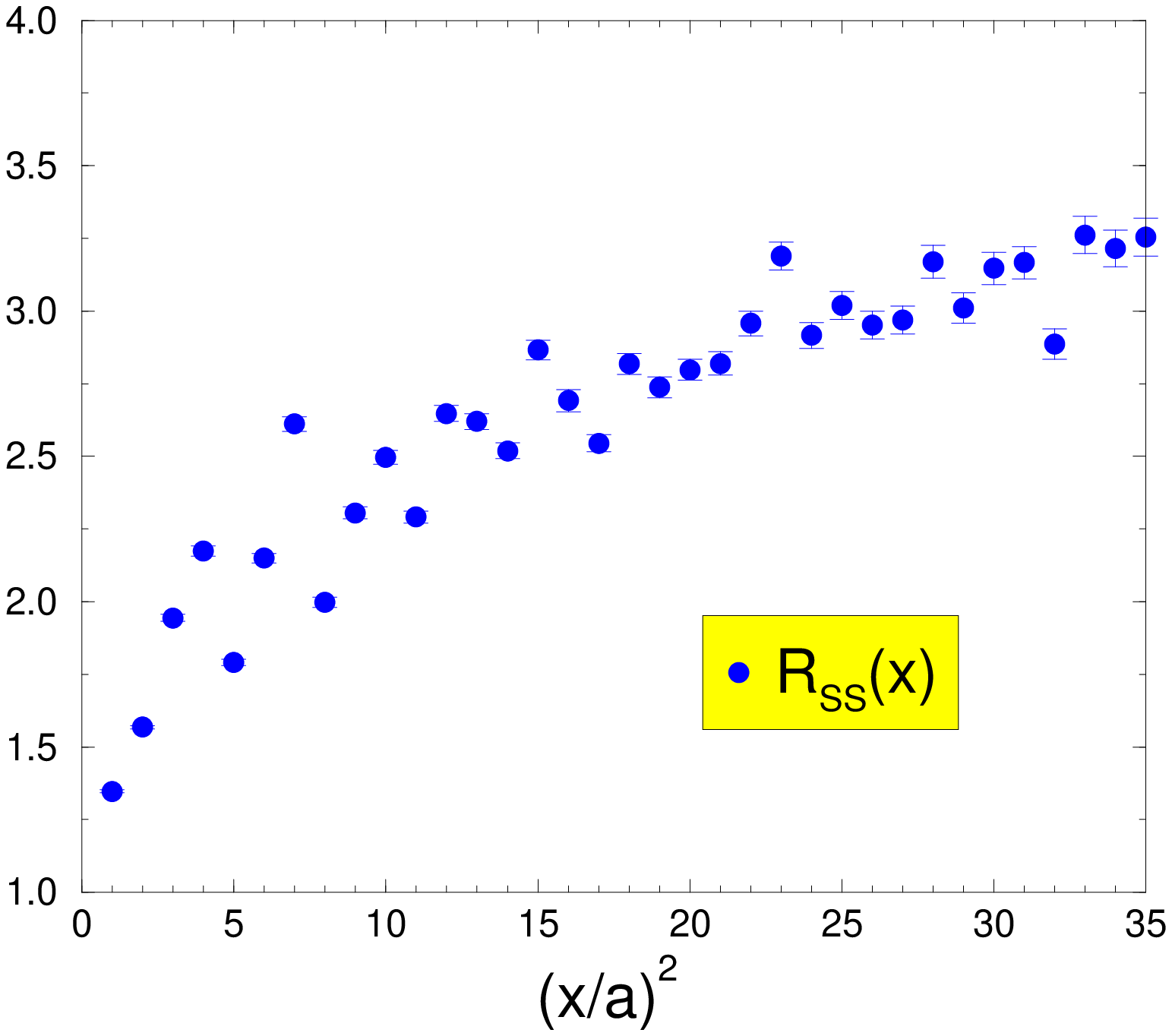} \\
\end{tabular}
\caption{\sl\small The ratios $R_{VV}(x)$ (left) and $R_{SS}(x)$ (right) 
defined in Eq.~(\ref{eq:rgamma}).\label{fig:vv_chi_int_rat}}
\vspace*{-0.5cm}
\end{figure}
Once the correlators have been corrected at tree level with the factor that
attempt to reduce discretization effects at the leading order, they still 
suffer for remaining discretization errors, at $\calo (g^2\, a^2)$ and higher. 
Their effects is shown in Fig.~\ref{fig:vv_chi_int_rat}, where we plot the 
corrected correlation functions $C_{\Gamma\Gamma}^{\prime}(x)$ normalized to 
their continuum counterparts in the free theory, 
\be
\label{eq:rgamma}
R_{\Gamma\Gamma}(x)=
\frac{C_{\Gamma\Gamma}^{\prime}(x)}{C_{\Gamma\Gamma}(x)_{\rm cont}^{\rm free}} 
\ee
in the case of the vector current and the scalar density operators. Note 
that, particularly in the scalar case, the remaining lattice artifacts are still
larger than statistical errors. A further reduction of these effects could 
be obtained by computing the $\calo (g^2\, a^2)$ terms in lattice perturbation 
theory.
 
In order to extract the renormalization constants, the previous analysis 
suggests to implement the renormalization condition defined 
in Eq.~(\ref{eq:rin}) directly to the corrected correlation functions 
$C_{\Gamma\Gamma}^{\prime}(x)$. This is equivalent to impose
\be\label{eq:rin_2}
\frac{\langle O^{\rm X}_\Gamma(x) O^{\rm X}_\Gamma(0) \rangle\;\;\;\;\;}
{\langle O_\Gamma(x) O_\Gamma(0) \rangle^{\rm free}_{\rm lat}} \Big
|_{x^2=x_0^2} = 1 \,.
\ee
Since $O^{\rm X}_\Gamma(x) = Z^{\rm X}_{\Gamma}(x_0) O_\Gamma (x)$, the above
condition implies
\be
Z^{\rm X}_{\Gamma}(x_0) = 1/\sqrt{R_{\Gamma\Gamma}(x_0)}\; .
\ee

\begin{figure}
\begin{tabular}{cc}
\epsfxsize7.5cm\epsffile{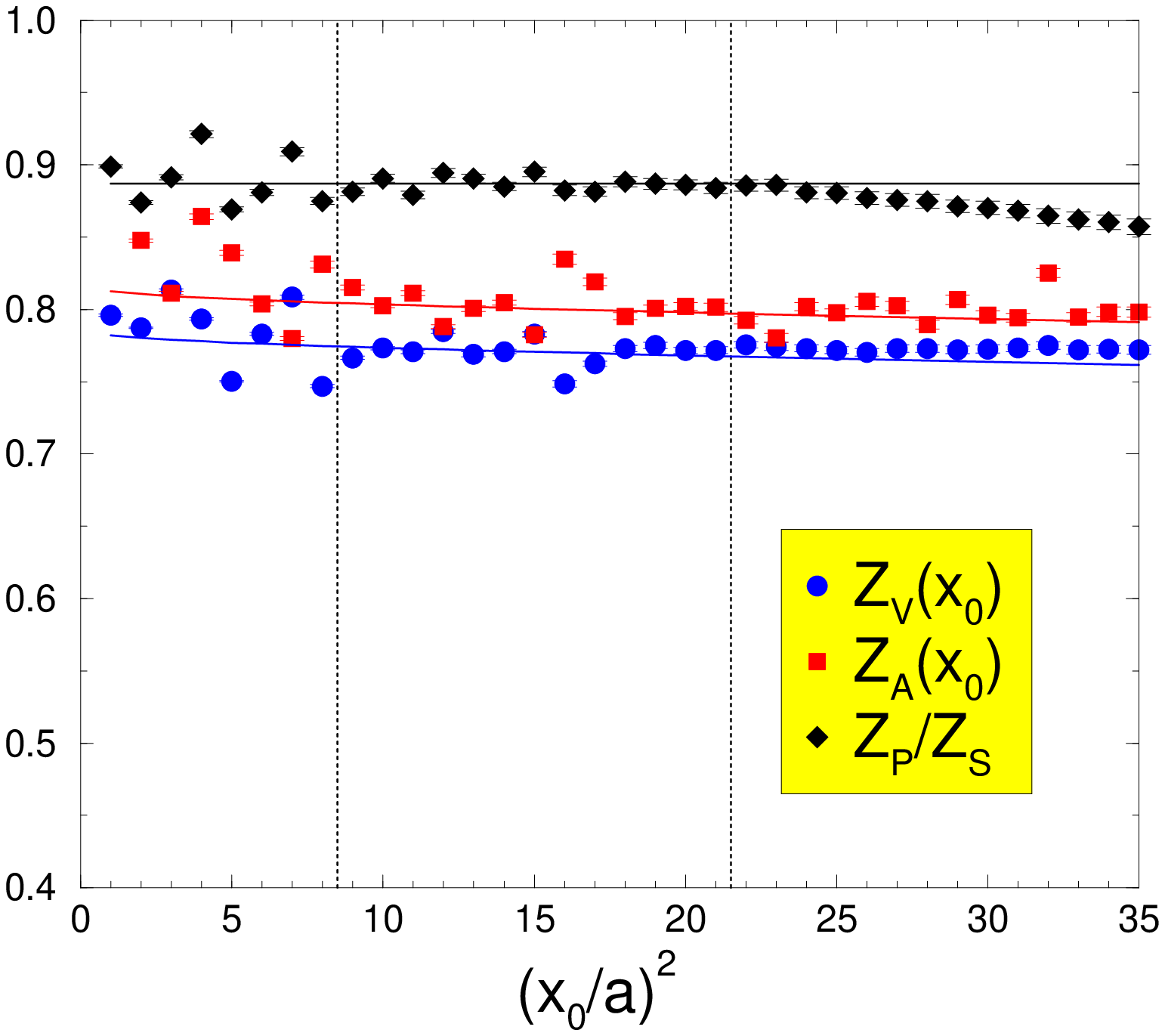} &
\hspace*{0.2cm}
\epsfxsize7.5cm\epsffile{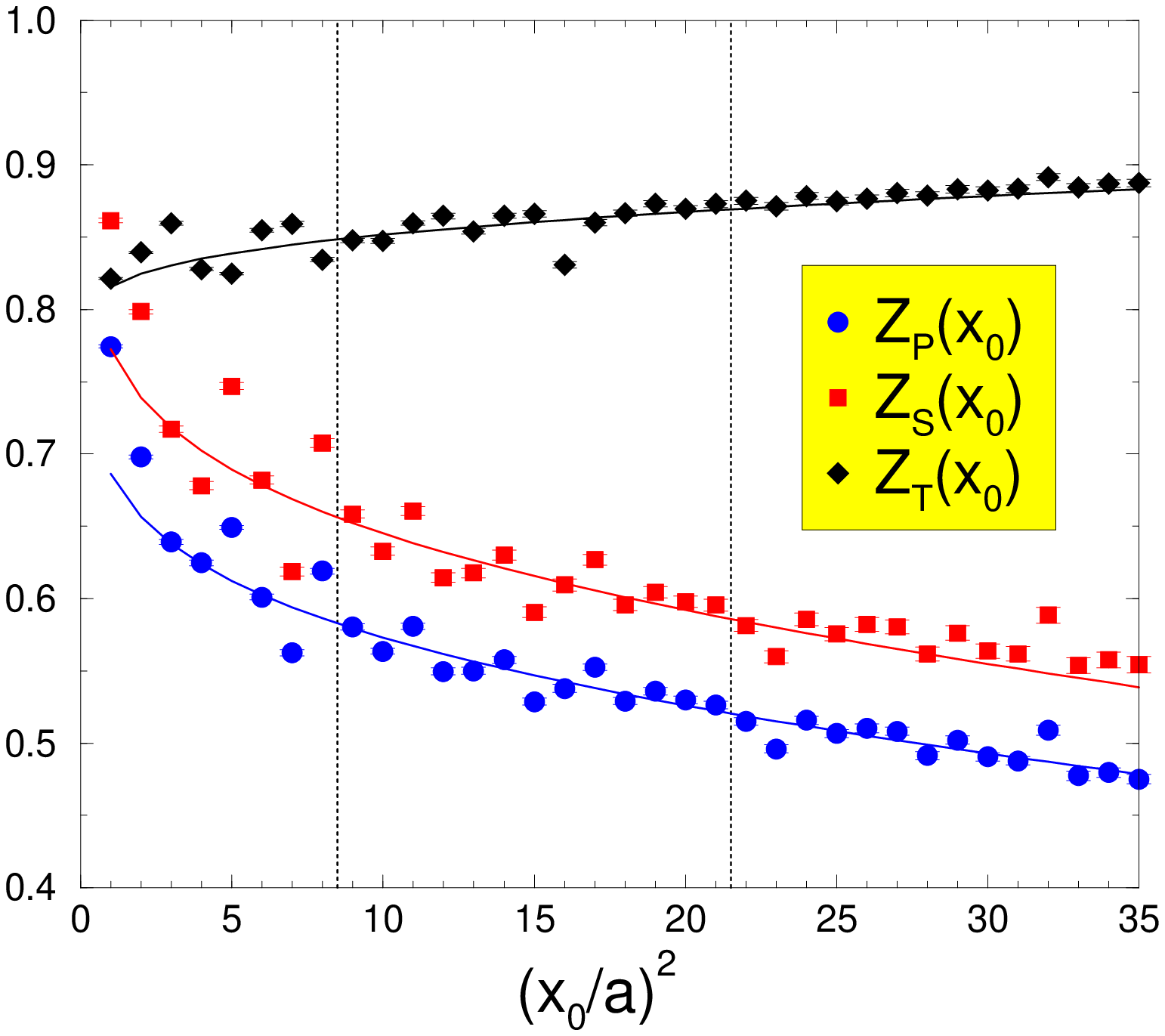} \\
\end{tabular}
\caption{\sl\small Renormalization constants in the X-space scheme as a function
of the renormalization scale $x_0$. The solid line indicate the scale dependence
predicted by NLO perturbation theory. The dashed lines show the boundaries 
of the fitting window \label{fig:vv_ss_chi_int}.}
\vspace*{-0.5cm}
\end{figure}
The renormalization constants $Z_V$, $Z_A$, $Z_S$, $Z_P$, $Z_T$ in the X-space
scheme and the ratio $Z_P/Z_S$ are shown in Fig.~\ref{fig:vv_ss_chi_int} as a 
function of $(x_0/a)^2$. The scattering is reduced for $(x_0/a)^2 \ge 8$,
signaling that discretization effects are moderate above this point. For 
$(x_0/a)^2 \le 21$, which corresponds to $1/x_0 \gtrsim 0.9$~GeV, the dependence
of the renormalization constants on the renormalization scale is compatible with
the NLO prediction of perturbation theory, indicated by solid lines in 
Fig.~\ref{fig:vv_ss_chi_int}. In particular, the ratio $Z_P/Z_S$ is expected to 
be independent of the renormalization scale and we see that in the range 
$(x_0/a)^2=[9,21]$ a reasonable good plateau is observed. In this window we 
extract the values of all renormalization constants by fitting the corresponding
correlation functions with the perturbative formul\ae~given in
sec.~\ref{sec:PT}. The results, translated to the $\MSbar$ scheme at the scale 
$\mu=2$ GeV are presented in Tab.~\ref{tab:results}.
\begin{table*}[t!]
\newcommand{\cc}[1]{\multicolumn{1}{c}{#1}}
\begin{center}
\begin{tabular}{clll}
 &\cc{$Z_V$}&\cc{$Z_A$}&\cc{$Z_T$}\\
\hline
This work&$0.801(2)(18)(6)$&$0.833(2)(27)(6)$&$0.895(2)(21)$\\
RI-MOM&$0.803(3)$&$0.833(3)$&$0.898(6)$\\
SF&$0.808(1)$&$0.825(8)$&$--$\\
\hline
\hline
\end{tabular}
\vskip 0.5cm
\begin{tabular}{clll}
 &\cc{$Z_P/Z_S$}&\cc{$Z_S$}&\cc{$Z_P$}\\
\hline
This work&$0.888(2)(8)$&$0.702(4)(27)(23)$&$0.624(3)(19)(21)$\\
RI-MOM&$0.897(4)$&$0.679(8)$&$0.609(8)$\\
SF&$0.912(9)$&$--$&$0.61(1)$\\
\hline
\hline
\end{tabular}
\end{center}
\caption{\sl\small Wilson results for the renormalization constants in the 
$\MSbar$ scheme at $\mu=2$ GeV, and comparison with other non-perturbative 
techniques: RI-MOM \cite{Becirevic:2004ny} and Schr\"odinger Functional 
\cite{Luscher:1996jn,Capitani:1998mq,Guagnelli:2000jw}.\label{tab:results}}
\vspace*{-0.5cm}
\end{table*}
The first quoted error is statistical and it is by far the smallest one. The 
second is an estimate of the uncertainty coming from the spread of the points 
within the fitting window. This error could be further reduced by going to 
finer lattices or by evaluating the $\calo (g^2\, a^2)$ terms in lattice 
perturbation theory. The third error in Tab.~\ref{tab:results} is an estimate of
the systematics due to higher orders in continuum perturbation theory, obtained 
by varying the renormalization scale $\mu$ in the perturbative expressions of 
Eq.~(\ref{eq:zz}) in the range $1\le \mu x_0 \le 2$. This uncertainty can be 
reduced by performing a N$^2$LO computation in perturbation theory and/or by 
implementing the step scaling technique\footnote{It is interesting to note that,
since the running of the operators with the renormalization scale is scheme 
dependent but regularization independent, it is possible to implement the step 
scaling technique by using any discretization of the fermionic action. The
results, extrapolated to the continuum limit, can then be used to evolve the 
renormalization constants computed non-perturbatively also with a different
discretization.} proposed in Ref.~\cite{Luscher:1991wu}. The latter would require
simulations at several lattice spacings and goes beyond the scope of this 
exploratory study.

\begin{figure}
\begin{center}
\begin{tabular}{c}
\epsfxsize7.5cm\epsffile{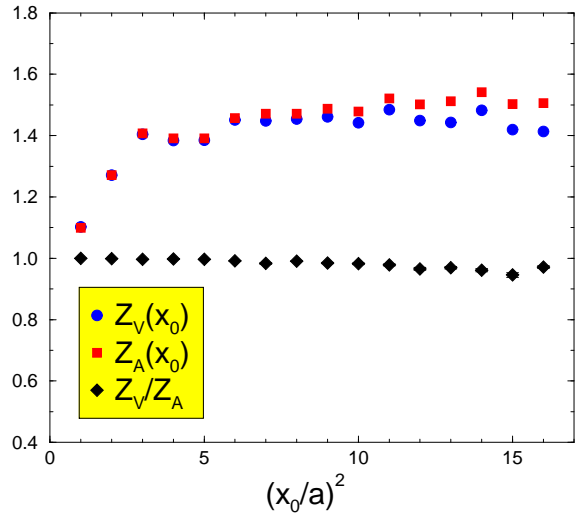} \\
\end{tabular}
\caption{\sl\small Renormalization constants $Z_V$, $Z_A$ and the ratio 
$Z_V/Z_A$ for Neuberger fermions as a function of $x_0$.
\label{fig:vv_ss_chi_int_ov}}
\end{center}
\vspace*{-0.5cm}
\end{figure}
In order to investigate the applicability of the X-space renormalization method
to different discretizations of the fermionic action, we also performed a
feasibility study by using Neuberger fermions. We used $80$ configurations 
generated with the same gluonic action at $\beta=6.0$ ($a\sim 0.093$~fm) and 
$V=16^3 \times 32$ which were retrieved from the repository at the 
``Gauge Connection'' \cite{GaugeConn}. For these configurations, overlap 
propagators at bare masses $ma=0.040$, 0.055, 0.070, 0.085, 0.100 have been 
evaluated, see Refs.~\cite{Giusti:2001pk,Giusti:2001yw} for details. The results
have been quadratically extrapolated to the chiral limit, as suggested by 
perturbation theory. The results for the renormalization constants of the vector
and axial-vector currents are shown in Fig.~\ref{fig:vv_ss_chi_int_ov} as a 
function of $(x_0/a)^2$. Since in this case the lattice spacing is larger than 
in the case of Wilson fermions, the window contains at most three points, which 
makes it difficult a reliable comparison with perturbation theory. Nevertheless 
the data plotted in Fig.~\ref{fig:vv_ss_chi_int_ov} show a smooth behaviour 
compatible with the chiral properties of Neuberger fermions, and the value of 
$Z_V=Z_A$ for $3 \le (x_0/a)^2\le 5$, corrected for the matching factor in 
Eq.~(\ref{eq:zz}), is compatible with the Ward identity estimate obtained in 
Ref.~\cite{Giusti:2001pk,Giusti:2001yw}, $Z_A = 1.55(4)$. In this case, the
implementation of a step scaling technique would have allowed us to impose the
renormalization conditions at shorter distances, where perturbation theory
is reliable but discretization effects remain negligible.

\section{Conclusions}\label{sec:concl}
We have studied the correlation functions of two fermion bilinear operators in 
coordinate space at short Euclidean distances. A good statistical signal can be 
obtained with a small number of configurations. The spread of the data computed 
at different lattice points with the same $x^2$, indicates that discretization 
errors can be large. This spread has been greatly reduced by normalizing the 
correlation functions with the analogous ones computed in the free theory at 
finite lattice spacing. A straightforward application of these results is a 
determination of the renormalization constants of the composite operators. Even 
if with larger uncertainties, the values of the renormalization constants which 
we have obtained are in good agreement with previous non-perturbative 
determinations. This technique can be easily applied to any fermion 
discretization, as shown in this paper, and it involves only gauge-invariant 
correlation functions among local operators at finite Euclidean distance. The
matching with more popular renormalization schemes, such as the $\MSbar$ scheme,
is easy because it requires calculations only performed in continuum 
perturbation theory. Therefore in the future, after more accurate studies of the
systematic, the X-space method could become a powerful technique to renormalize 
the four-fermion operators relevant in hadronic weak decays. 

\section*{Acknowledgments}
\indent It is a pleasure to thank C.~Hoelbling, C.~Rebbi and M.~Testa for 
stimulating discussions. Thanks also to M.~L\"uscher and R.~Sommer for 
useful comments on the paper. We also thank C.~Hoelbling and C.~Rebbi for allowing 
us to use the quark propagators generated with the Neuberger action in 
Ref.~\cite{Giusti:2001pk,Giusti:2001yw}. We gratefully acknowledge the use of 
the gauge configurations produced by the authors in Ref.~\cite{GaugeConn}. S.P.
work was partially supported by the M.U.R.S.T. This work has been funded by 
MCyT, Plan Nacional I+D+I (Spain) under the Grant BFM2002-00568.


\begin{thebibliography}{99}

\bibitem{Martinelli:1994ty}
G.~Martinelli, C.~Pittori, C.~T.~Sachrajda, M.~Testa and A.~Vladikas,
Nucl.\ Phys.\ B {\bf 445} (1995) 81, [hep-lat/9411010].

\bibitem{Gimenez:1998ue}
V.~Gimenez, L.~Giusti, F.~Rapuano and M.~Talevi,
Nucl.\ Phys.\ B {\bf 531} (1998) 429, [hep-lat/9806006].

\bibitem{Gockeler:1998ye}
M.~Gockeler {\it et al.},
Nucl.\ Phys.\ B {\bf 544} (1999) 699, [hep-lat/9807044].

\bibitem{Donini:1999sf}
A.~Donini, V.~Gimenez, G.~Martinelli, M.~Talevi and A.~Vladikas,
Eur.\ Phys.\ J.\ C {\bf 10} (1999) 121, [hep-lat/9902030].

\bibitem{Giusti:2000jr}
L.~Giusti and A.~Vladikas,
Phys.\ Lett.\ B {\bf 488} (2000) 303, [hep-lat/0005026].

\bibitem{Blum:2001sr}
T.~Blum {\it et al.},
Phys.\ Rev.\ D {\bf 66} (2002) 014504, [hep-lat/0102005].

\bibitem{Giusti:2001pk}
L.~Giusti, C.~Hoelbling and C.~Rebbi,
Phys.\ Rev.\ D {\bf 64} (2001) 114508
[Erratum-ibid.\ D {\bf 65} (2002) 079903], [hep-lat/0108007].

\bibitem{Giusti:2001yw}
L.~Giusti, C.~Hoelbling and C.~Rebbi,
Nucl.\ Phys.\ Proc.\ Suppl.\  {\bf 106} (2002) 739, [hep-lat/0110184].

\bibitem{Becirevic:2004ny}
D.~Becirevic, V.~Gimenez, V.~Lubicz, G.~Martinelli, M.~Papinutto and J.~Reyes,
arXiv:hep-lat/0401033.

\bibitem{Gattringer:2004iv}
C.~Gattringer, M.~Gockeler, P.~Huber and C.~B.~Lang,
arXiv:hep-lat/0404006.


\bibitem{Luscher:1992an}
M.~Luscher, R.~Narayanan, P.~Weisz and U.~Wolff,
Nucl.\ Phys.\ B {\bf 384} (1992) 168, [hep-lat/9207009].

\bibitem{Luscher:1991wu}
M.~Luscher, P.~Weisz and U.~Wolff,
Nucl.\ Phys.\ B {\bf 359} (1991) 221.

\bibitem{Luscher:1996jn}
M.~Luscher, S.~Sint, R.~Sommer and H.~Wittig,
Nucl.\ Phys.\ B {\bf 491} (1997) 344,
[hep-lat/9611015].

\bibitem{Capitani:1998mq}
S.~Capitani, M.~Luscher, R.~Sommer and H.~Wittig  [ALPHA Collaboration],
Nucl.\ Phys.\ B {\bf 544} (1999) 669, [hep-lat/9810063].

\bibitem{Luscher:1982uv}
M.~Luscher,
Phys.\ Lett.\ B {\bf 118} (1982) 391.

\bibitem{Luscher:1982ma}
M.~Luscher,
Nucl.\ Phys.\ B {\bf 219} (1983) 233.

\bibitem{Martinelli:1997zc}
G.~Martinelli et al.,
Phys.\ Lett.\ B {\bf 411} (1997) 141, [hep-lat/9705018].

\bibitem{Becirevic:2002yv}
D.~Becirevic et al., 
Nucl.\ Phys.\ Proc.\ Suppl.\  {\bf 119} (2003) 442, [hep-lat/0209168].

\bibitem{Gimenez:2003rt}
V.~Gimenez {\it et al.},
arXiv:hep-lat/0309150.

\bibitem{Chetyrkin:1996sr}
K.~G.~Chetyrkin,
Phys.\ Lett.\ B {\bf 390} (1997) 309, [arXiv:hep-ph/9608318].

\bibitem{Gorishnii:1990vf}
S.~G.~Gorishnii, A.~L.~Kataev and S.~A.~Larin,
Phys.\ Lett.\ B {\bf 259} (1991) 144.

\bibitem{Surguladze:1990tg}
L.~R.~Surguladze and M.~A.~Samuel,
Phys.\ Rev.\ Lett.\  {\bf 66} (1991) 560, [Erratum-ibid.\  {\bf 66} (1991) 2416].



\bibitem{Bochicchio:1985xa}
M.~Bochicchio, L.~Maiani, G.~Martinelli, G.~C.~Rossi and M.~Testa,
Nucl.\ Phys.\ B {\bf 262} (1985) 331.

\bibitem{Chetyrkin:1997dh}
K.~G.~Chetyrkin,
Phys.\ Lett.\ B {\bf 404} (1997) 161, [hep-ph/9703278].

\bibitem{Vermaseren:1997fq}
J.~A.~M.~Vermaseren, S.~A.~Larin and T.~van Ritbergen,
Phys.\ Lett.\ B {\bf 405} (1997) 327, [hep-ph/9703284].

\bibitem{Gracey:2000am}
J.~A.~Gracey,
Phys.\ Lett.\ B {\bf 488} (2000) 175, [hep-ph/0007171].

\bibitem{GaugeConn}
We retrieved the first $80$ gauge configurations OSU\_Q60a produced by
G.~Kilcup, D.~Pekurovsky, L.~Venkataraman, Nucl. Phys. (Proc. Suppl.)  53 (1997)
345 from the repository at the ``Gauge Connection'' (http://qcd.nersc.gov/). 

\bibitem{Guagnelli:2000jw}
M.~Guagnelli, R.~Petronzio, J.~Rolf, S.~Sint, R.~Sommer and U.~Wolff  [ALPHA
                  Collaboration],
Nucl.\ Phys.\ B {\bf 595} (2001) 44, [hep-lat/0009021].
\end{thebibliography}
\end{document}